\documentclass[12pt]{article}
\usepackage[T2A]{fontenc}
  \usepackage[english]{babel}

\usepackage{latexsym}
\usepackage{amsmath}
\usepackage{amssymb}
\usepackage{bm}
\usepackage{graphicx}
\usepackage{indentfirst}
\usepackage{cite}

\newcommand{\be}{\begin{equation}}
\newcommand{\ee}{\end{equation}}

\newcommand{\bea}{\begin{eqnarray}}
\newcommand{\eea}{\end{eqnarray}}

\usepackage[dvips]{color}

\date{\today}

\title{Non-paraxial relativistic wave packets \\ with orbital angular momentum}

\author{Dmitry Karlovets}
\date{{\small Tomsk State University, Lenina 36, 634050 Tomsk, Russia}}

\begin{document}

\maketitle

\begin{abstract}
One of the reasons for the tremendous success of a plane-wave approximation in particle physics is that the non-paraxial corrections to such observables as
energy, magnetic moment, scattering cross section, and so on are attenuated as $\lambda_c^2/\sigma_{\perp}^2 \ll 1$ where $\sigma_{\perp}$ is a beam width 
and $\lambda_c = \hbar/mc$ is a Compton wavelength. This amounts to less than $10^{-14}$ for modern electron accelerators
and less than $10^{-6}$ for electron microscopes. Here we show that these corrections are $|\ell|$ times enhanced
for vortex particles with high orbital angular momenta $|\ell|\hbar$, which can already be as large as $10^3\hbar$.
We put forward the relativistic wave packets, both for vortex bosons and fermions, which transform correctly under the Lorentz boosts, are localized in a 3D space, 
and represent a non-paraxial generalization of the Laguerre-Gaussian beams. We demonstrate that it is $\sqrt{|\ell|}\, \lambda_c \gg \lambda_c$
that defines a paraxial scale for such packets, in contrast to those with a non-singular phase (say, the Airy beams). 
With current technology, the non-paraxial corrections can reach the relative values of $10^{-3}$, yield a proportional increase of an invariant mass of the electron packet, 
describe a spin-orbit coupling as well as the quantum coherence phenomena in particle and atomic collisions.
\end{abstract}

\section{Introduction}

Particles carrying orbital angular momentum (OAM) can be described with the non-localized Bessel states (akin to a plane wave) \cite{Bliokh, Review}.
Thanks to a finite transverse momentum, they predict a spin-orbit interaction and other non-paraxial phenomena. 
Nevertheless these states can be unsuitable for quantitative description of the experiments with tightly focused vortex beams if their finite width, length, and monochromaticity are of importance.
This can be the case, for instance, when the beams are focused to a spot with a size $\sigma_{\perp}$ comparable to a characteristic scale of a problem, which is a Bohr radius $a \approx 0.053$ nm for atomic physics and an electron's Compton wavelength $\lambda_c = \hbar/mc \approx 0.39$ pm for particle physics. With the current record of $\sigma_{\perp} \gtrsim 0.1$ nm $\approx 2a$ \cite{Angstrom},
a more realistic wave-packet treatment beyond the paraxial approximation is needed, in particular, for proper study of the spin-orbit phenomena
and for scattering problems in atomic and high-energy physics, especially when the quantum interference and coherence play a notable role \cite{Sarkadi, Sch, PRL}.
The quantum interference can be of crucial importance -- say, for potential applications in hadronic physics \cite{JHEP} -- and the orthogonal Bessel states are not applicable here either. 

In collisions of the Gaussian packets, non-paraxial corrections to a scattering cross section are generally attenuated as $\lambda_c^2/\sigma_{\perp}^2 \ll 1$ and do not play any essential role.
Dealing with the focused vortex packets instead, these corrections can be enhanced when the OAM is large, $|\ell|\hbar \gg \hbar$ \cite{JHEP} 
(it can be as high as $|\ell| \sim 10^3$ \cite{l1000}). Precise estimates of these phenomena also require that the vortex packets be spatially localized, described in a Lorentz invariant way, 
and applicable beyond the paraxial approximation. Despite the recent interest in the relativistic packets with OAM \cite{Barnett, Birula}, 
such a model is still lacking (see the recent discussion in \cite{Review,Bliokh17}) first and foremost because the widely used coordinate representation is inconvenient for these purposes.
% The situation here is somewhat reminiscent of that with a photon wave function in x-space \cite{Birula_photon}.

A covariant description of the relativistic wave packets has been recently given in \cite{Packets1, Packets2, Naumov} for massive neutrinos. 
In a widespread model, the packet is Gaussian in $p$-space, it has a (mean) 4-momentum $\bar{p}_{\mu}$, $\bar{p}^2 = m^2$, and a scalar uncertainty $\sigma$, 
which is vanishing, $\sigma \ll m$, in the paraxial regime. In this paper we further develop this approach  by adding OAM to the set of quantum numbers, both for bosons and fermions.
Such vortex packets are localized in a 3D space, transform correctly under the Lorentz boosts and, being exact solutions to the relativistic wave equations,
represent a non-paraxial generalization of the Laguerre-Gaussian (LG) beams.
%most importantly, can be used for quantitative estimates of the non-paraxial phenomena.

We calculate the electron's mean energy, magnetic moment, etc., and demonstrate that the non-paraxial effects are $|\ell|$ times \textit{enhanced} for vortex packets 
compared to the Gaussian beam or even to that with a non-singular phase (say, an Airy packet \cite{Airy, Airy_El_Exp}). 
We argue that this enhancement \textit{cannot be reproduced} with the known Bessel or LG states.
For instance, correction to the electron invariant mass is found to be positive and for available beams it can reach the values of $(10^{-4}-10^{-3})\,m$. 
To put it differently, such a wave packet is $0.01\%-0.1\%$ heavier than an ordinary plane-wave electron. 
It is this weighting that can reveal itself in the corresponding corrections, 
$$
\sim |\ell| \lambda_c^2/\sigma_{\perp}^2 \gtrsim \alpha_{em}^2 = 1/137^2 \gg \lambda_c^2/\sigma_{\perp}^2,
$$ 
to the $e^-e^-$, $e^-e^+$ or $e^-\gamma$ high-energy scattering, which can compete with the two-loop QED contributions.

Thus, sub-nm-sized highly twisted beams can become a useful tool for probing the previously unexplored non-paraxial phenomena in high-energy and nuclear physics,
analogously to the quantum coherence effects in atomic collisions \cite{Sarkadi, Sch} and in addition to the magnetic-moment phenomena in radiation \cite{PRL13}. 
Moreover, in scattering of a superposition $|\ell_1\rangle + |\ell_2\rangle$ by atoms the analogous fundamental scale, the Bohr radius $a$, is $1/\alpha_{em} = 137$ times larger than $\lambda_c$. 
As a result, the corresponding effects may become only moderately attenuated (akin to \cite{PRL}), and one would need an explicitly non-paraxial approach, 
feasible with the packets presented in this paper. A system of units $\hbar = c = e = 1$ is used and the metric is $g_{\mu\nu} = \text{diag}(1,-1,-1,-1)$.

\section{Relativistic scalar packets with OAM}

A wave function $\psi_{\ell}(p)$ of a packet with OAM $\ell$ is to be Lorentz invariant for longitudinal boosts and, 
therefore, it can depend only on a scalar $(p_{\mu} - \bar{p}_{\mu})^2 \leq 0$. Such an OAM-orthogonal and normalized function is 
\begin{eqnarray}
& \displaystyle \psi_{\ell}(p) = \frac{2^{3/2} \pi}{\sigma^{|\ell| + 1}\sqrt{|\ell|!}}\, p_{\perp}^{|\ell|}\,\frac{e^{-m^2/\sigma^2}}{\sqrt{K_{|\ell| + 1} (2m^2/\sigma^2)}}
 \exp\left\{\frac{(p_{\mu} - \bar{p}_{\mu})^2}{2\sigma^2} + i\ell\phi_p\right\},\cr
& \displaystyle \int \frac{d^3p}{(2\pi)^3}\frac{1}{2\varepsilon}\,\left[\psi_{\ell^{\prime}} (p)\right]^* \psi_{\ell} (p) = \delta_{\ell,\ell^{\prime}},
\label{OAMrelp}
\end{eqnarray}
where $\bar{p}^{\mu} = \{\bar{\varepsilon}, 0, 0,\bar{p}\}, \bar{\varepsilon} = \sqrt{\bar{p}^2 + m^2}, p^2 = \bar{p}^2 = m^2, \varepsilon = \sqrt{{\bm p}^2 + m^2}$, ${\bm p} = \{{\bm p}_{\perp}, p_z\}$, and $K_{|\ell| + 1}$ 
is a modified Bessel function. Due to azimuthal symmetry of the packet, we restrict ourselves to the longitudinal boosts; see \cite{Bliokh17trans} for effects of the transverse ones.

A Fourier transform of this function,
\begin{eqnarray}
& \displaystyle \psi_{\ell}(x) =  \int \frac{d^3 p}{(2\pi)^3} \frac{1}{2\varepsilon}\, \psi_{\ell}(p) e^{-ipx} = \frac{(i\rho)^{|\ell|}}{\sqrt{2|\ell|!}\, \pi}\,\frac{\sigma^{|\ell| + 1}}{\varsigma^{|\ell|+1}}\,\frac{K_{|\ell|+1} (\varsigma m^2/\sigma^2)}{\sqrt{K_{|\ell|+1}(2m^2/\sigma^2)}}\,
e^{i\ell\phi_r},\cr
& \displaystyle \varsigma = \frac{1}{m}\sqrt{\left (\bar{p}_{\mu} + ix_{\mu} \sigma^2\right )^2} = \text{inv},\, \text{Re}\,\varsigma > 0,
\label{OAMrelexact}
\end{eqnarray}
represents an exact solution to the Klein-Gordon equation. Here $x^{\mu} = \{t, {\bm r}\}, {\bm r} = \{{\bm \rho}, z\}$. Clearly, for such a packet $\langle \hat{L}_z\rangle = \ell$ where $\hat{L}_z = -i\partial/\partial\phi_p$ or $\hat{L}_z = -i\partial/\partial\phi_r$, depending on the representation. 

As we show in details elsewhere \cite{PRA2018}, in the paraxial regime the function (\ref{OAMrelexact}) reduces to the invariant LG beam with a radial index $n=0$ 
(generalization for $n\ne 0$ is straightforward but not relevant for current purposes):
\begin{eqnarray}
& \displaystyle \psi^{\text{par}}_{\ell}(x) =  \frac{i^{\ell}}{\sqrt{|\ell|!}}\frac{1}{\sqrt{2m}}\left(\frac{\sigma}{\sqrt{\pi}}\right)^{3/2}\frac{(\rho/\sigma_{\perp}(t))^{|\ell|}}{(1+(t/t_d)^2)^{3/4}} 
\exp\Big\{i\ell\phi_r - i\bar{p}^{\mu}x_{\mu} - \cr
& \displaystyle - i \left(|\ell| + 3/2\right) \arctan \left(t/t_d\right) - \frac{1}{2\sigma^2_{\perp}(t)}\Big(1-it/t_d\Big)\Big(\rho^2 + \frac{\bar{\varepsilon}^2}{m^2}(z-\bar{u}t)^2\Big)\Big\},
\label{LG}
\end{eqnarray}
where $t_d = \bar{\varepsilon}/\sigma^2$ is a diffraction time, $\sigma_{\perp}(t) = \sigma^{-1}\sqrt{1 + (t/t_d)^2}$ is the beam width, and $\bar{u} = \bar{p}/\bar{\varepsilon}$. 
Note that $\sigma_{\perp}(t)$, $t/t_d$, and $\rho^2 + \bar{\varepsilon}^2(z-\bar{u}t)^2/m^2$ are Lorentz invariant, the Gouy phase has $3/2$ instead of $1 \equiv 2/2$ because of the packet's 3D localization in space and, in contrast to Eq.(12) in \cite{Barnett}, it also depends on $t/t_d$ instead of $z/z_R$ 
because of massiveness of the packet. As can be seen from (\ref{LG}), it is only for relativistic energies, $\bar{\varepsilon} \gg m$, that one can substitute $t/t_d \rightarrow z/z_R$
and hardly for 200-300 keV beams of the electron microscopes. This is because the condition of paraxiality, $\sigma \ll m$, under which Eq.(\ref{OAMrelexact}) reduces to Eq.(\ref{LG}) is Lorentz invariant,
while the usually used one, $p_{\perp} \ll p_z$, borrowed from optics with massless photons, is not so. Hence the resultant non-invariant LG beams with $z/z_R$ (say, those of \cite{Barnett}) 
are applicable only for relativistic energies.
%The transverse probability density represents \textit{a gamma distribution} (or a Poisson distribution for $|\ell|$; see, for instance, \cite{Mandel}),
%\begin{eqnarray}
%& \displaystyle
%|\psi_{\ell}^{\text{par}}({\bm \rho}, z = t = 0)|^2 =\text{const}\, \frac{(\rho \sigma)^{2|\ell|}}{|\ell|!}\, e^{-(\rho\sigma)^2},
%\label{j0paraxx}
%\end{eqnarray} 
%with its typical for a vortex state doughnut-like profile with a maximum at $(\rho \sigma)^2 = |\ell|$. 

Regardless of the OAM, the exact wave function (\ref{OAMrelexact}) decays exponentially at large distances, $\sqrt{-x_{\mu}^2} \gg \lambda_c$, 
\begin{eqnarray}
& \displaystyle \psi_{\ell}(x) \propto \exp\left\{-\sqrt{-x_{\mu}^2}/\lambda_c\right\}.
\label{psiass}
\end{eqnarray}
Within this class of functions, it is the only law allowed by the invariance considerations and it is in sharp contrast with Eq.(15) in Ref.\cite{Birula} 
where the analogous scale is $\bar{\varepsilon}/m$ times smaller than $\lambda_c$, which is impossible within a one-particle approach with a stable vacuum (see, for instance, \cite{BLP}) 
and is not Lorentz invariant. 

One can calculate the mean energy and momentum exactly by using the energy-momentum tensor $T_{\mu\nu}$ \cite{BLP},
\begin{eqnarray}
& \displaystyle \langle p^{\mu}_{\ell}\rangle = \{\langle \varepsilon_{\ell}\rangle,\langle {\bm p}^i_{\ell}\rangle\} = \int d^3r \{T^{00}, {\bm T}^{i0}\} = \cr
& \displaystyle = \int \frac{d^3p}{(2\pi)^3} \frac{|\psi_{\ell}(p)|^2}{2\varepsilon} \{\varepsilon, {\bm p}^i\} = 
\{\bar{\varepsilon},\bar{\bm p}^i\}\, \frac{K_{|\ell|+2}\left(2m^2/\sigma^2\right)}{K_{|\ell|+1}\left(2m^2/\sigma^2\right)}.
\label{energymeanexvortex}
\end{eqnarray}
Now we demand that $\sigma/m \equiv \lambda_c/\sigma_{\perp}$ be a small parameter and expand the Bessel functions in series. 
This yields a non-paraxial correction,
\begin{eqnarray}
& \displaystyle \langle p^{\mu}_{\ell}\rangle \simeq \{\bar{\varepsilon},\bar{\bm p}\}\,\left(1 + \left(\frac{3}{4} + \frac{|\ell|}{2}\right) \frac{\sigma^2}{m^2}\right).
\label{energymeanexvortex2}
\end{eqnarray}
It is always positive and it depends on the OAM, which can be arbitrarily large (provided that the correction is still small). 
As we shall see, the same expression also holds for a fermion, that is why these terms, 
\begin{eqnarray}
& \displaystyle |\ell| \frac{\lambda_c^2}{\sigma_{\perp}^2} \gg \frac{\lambda_c^2}{\sigma_{\perp}^2} = \mathcal O(\hbar^2)
\label{corr}
\end{eqnarray}
describe quantum corrections to the packet motion and they have no relation to the so-called Zitterbewegung.
	
An invariant mass of the packet,
\begin{eqnarray}
& \displaystyle m_{\ell}^2 = \langle p_{\ell}\rangle^2 \simeq m^2\,\left(1 + \left(\frac{3}{2} + |\ell|\right) \frac{\sigma^2}{m^2}\right),
\label{invmassvortex}
\end{eqnarray}
also depends on $|\ell|$. One can say that a vortex packet is heavier than a plane-wave electron,
\begin{eqnarray}
& \displaystyle \frac{\delta m_{\ell}}{m} \equiv \frac{m_{\ell} - m_{\ell=0}}{m} = \frac{|\ell|}{2} \frac{\sigma^2}{m^2} + \mathcal O (\sigma^4/m^4).
\label{deltam}
\end{eqnarray}
For beams with $|\ell| \gtrsim 10^3$ focused to a spot of $\sigma_{\perp} < 1\,\text{nm}$, we have
$$
\frac{\delta m_{\ell}}{m} < 10^{-3}.
$$

Thus, the non-paraxial corrections are $|\ell|$ times \textit{enhanced} for highly twisted packets. 
To put it differently, it is no longer the Compton wavelength that defines a paraxial scale, but it is
$$
\sqrt{\ell}\,\lambda_c,
$$
which can be more than an order of magnitude larger than $\lambda_c$ for available beams. 

\begin{figure}[t]
	%\hspace*{-0.2cm}
	\center
	\includegraphics[width=0.55\linewidth]{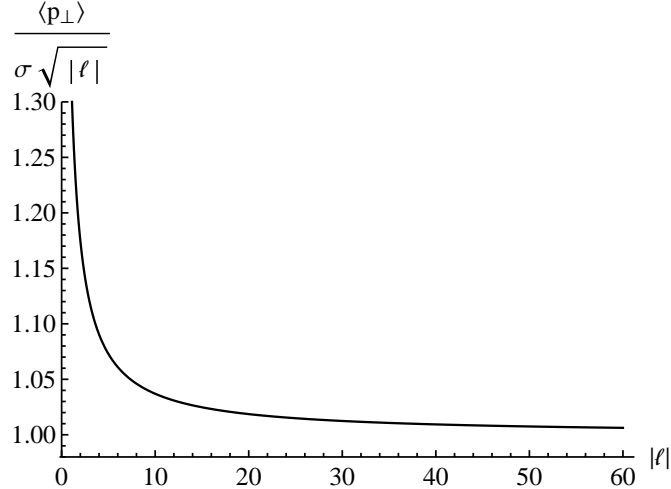}
	\caption{The mean transverse momentum of the vortex wave packet (\ref{OAMrelp}) in the paraxial regime, $|\ell|\sigma^2/m^2 \ll 1$.
\label{Figell}}
\end{figure}

Similarly, other observables and their non-paraxial corrections can also depend on $|\ell|$. For instance, one can calculate the mean transverse momentum exactly,
\begin{eqnarray}
& \displaystyle \langle p_{\perp}\rangle = \sigma \frac{\Gamma (|\ell| + 3/2)}{\Gamma (|\ell| + 1)}\,\frac{K_{|\ell| + 3/2}(2m^2/\sigma^2)}{K_{|\ell| + 1}(2m^2/\sigma^2)} \approx \cr
& \displaystyle \approx \sigma \sqrt{|\ell|} \left(1 + \mathcal O (|\ell|\sigma^2/m^2)\right)\, \text{when}\,\, |\ell| \gg 1.
\label{pperpmean}
\end{eqnarray}
This $\sqrt{|\ell|}$ dependence in the leading order is illustrated in Fig.\ref{Figell}. % Say, for $|\ell| \sim 10^3$ we have $\langle p_{\perp}\rangle \simeq \sqrt{10^3}\sigma \sim 30\sigma$. 
The invariant condition of paraxiality can now be rewritten as follows:
\begin{eqnarray}
& \displaystyle |\ell|\frac{\sigma^2}{m^2} \simeq \frac{\langle p_{\perp}\rangle^2}{m^2} = \left(\frac{\bar{\varepsilon}^2}{m^2} - 1\right) \tan^2\theta_0 \ll 1,
\label{ineqnonpar}
\end{eqnarray}
where an opening angle $\theta_0 = \arctan \langle p_{\perp}\rangle/\bar{p}$, in contrast to the Bessel beam, \textit{grows with} $|\ell|$. 
One can say that enhancement of the non-paraxial effects for vortex beams owes to their large transverse momentum or to the large angle $\theta_0$.
We would like to emphasize that these results cannot be obtained with the Bessel beams or the LG packets,
as the former have a definite transverse momentum $\varkappa$, which is independent of $\ell$, while the latter yield just $\langle p_{\perp}\rangle = \sigma \sqrt{|\ell|}$ without the non-paraxial correction.

\section{Relativistic fermion packets with OAM}

One can define a wave function for a fermion packet as follows:
\begin{eqnarray}
\psi_f (p) = \frac{u(p)}{\sqrt{2\varepsilon}}\, \psi(p),\ \int \frac{d^3p}{(2\pi)^3}\frac{1}{2\varepsilon}\,|\psi_f (p)|^2 = 1 = \text{inv},
\label{WFf}
\end{eqnarray}
where $\psi(p)$ is a normalized bosonic wave function (say, (\ref{OAMrelp})) and 
$$
u(p) = \left (\sqrt{\varepsilon + m}\,\omega, \sqrt{\varepsilon - m}\,({\bf p}{\bm \sigma})\omega/|{\bf p}|\right )^{T}
$$
is a bispinor, which obeys $|u(p)|^2 = 2\varepsilon$. In what follows, we deal with the helicity states for which a 2-component spinor $\omega$ obeys
\begin{eqnarray}
& \displaystyle 
(\hat{\bm z}{\bm \sigma})\, \omega = 2\lambda\, \omega,\ \hat{\bm z} = \bar{\bf p}/|\bar{\bf p}|,\, \lambda = \pm 1/2,\, \omega^{\dagger}\omega = 1,
\label{lambda}
\end{eqnarray}
where ${\bm \sigma}$ are the Pauli matrices. Note that we project the spin onto the packet's ``mean momentum'' $\bar{\bf p}$ and not onto ${\bf p}$, which is an integration variable.

A wave function in x-space, which is an exact solution to the Dirac equation, can be defined as follows:
\begin{eqnarray}
& \displaystyle \psi_f (x) = \int\frac{d^3p}{(2\pi)^3}\frac{1}{\sqrt{2\varepsilon}}\, \psi_f (p) e^{-ipx}. 
% = \cr & \displaystyle \qquad \qquad = \int\frac{d^3p}{(2\pi)^3}\,\frac{u(p)}{2\varepsilon}\, \psi (p) e^{-ipx}.
\label{WFfx}
\end{eqnarray}
For the packets studied in this paper, the integral can be evaluated via the steepest descent method. The resultant paraxial function, however, is of little use
and it is more convenient to work in p-space. The 4-current is $j = \{|\psi_f (x)|^2, \bar{\psi}_f(x){\bm \gamma}\psi_f(x)\}$
and the normalization $\int d^3r j^0 = 1$ coincides with that in (\ref{WFf}).

When $\psi(p)$ is from (\ref{OAMrelp}), the function $\psi_f (x)$ describes a vortex electron packet
with a total angular momentum 
$$
\langle\hat{j_z}\rangle = \ell + \lambda,
$$ 
as can be easily shown by acting by an operator $\hat{j_z} = \hat{L}_z + \hat{s}_z$ (here $\hat{s}_z = 1/2\,\text{diag}(\sigma_3, \sigma_3)$\cite{BLP}).
Clearly, the mean energy and momentum of this state coincide with those in the spinless formula (\ref{energymeanexvortex}).

We now turn to calculations of the intrinsic magnetic moment, which for an arbitrary packet is
%\begin{widetext}
\begin{eqnarray}
& \displaystyle 
{\bm \mu}_f = \frac{1}{2} \int d^3r\, {\bf r} \times \bar{\psi}_f(x){\bm \gamma}\psi_f(x) = \cr
& \displaystyle  = \frac{1}{2} \int \frac{d^3p}{(2\pi)^3}\frac{d^3k}{(2\pi)^3}\,d^3r\,\frac{\psi^*({\bf p} - {\bf k}/2) \psi({\bf p} + {\bf k}/2)}{2\varepsilon({\bf p} - {\bf k}/2) 2\varepsilon({\bf p} + {\bf k}/2)}\,\, {\bf r}\times \bar{u}({\bf p} - {\bf k}/2){\bm \gamma} u({\bf p} + {\bf k}/2)\cr
& \displaystyle \times \exp\left\{-it[\varepsilon ({\bf p} + {\bf k}/2) - \varepsilon ({\bf p} - {\bf k}/2)] + i{\bf r} {\bf k}\right\} = \cr
& \displaystyle  = \frac{1}{2} \int \frac{d^3p}{(2\pi)^3}\,d^3k\, \delta({\bf k})\,\, i\frac{\partial}{\partial {\bf k}} \times \bar{u}({\bf p} - {\bf k}/2){\bm \gamma} u({\bf p} + {\bf k}/2)\,\,
\frac{\psi^*({\bf p} - {\bf k}/2) \psi({\bf p} + {\bf k}/2)}{2\varepsilon({\bf p} - {\bf k}/2) 2\varepsilon({\bf p} + {\bf k}/2)},
\label{muf}
\end{eqnarray}
%\end{widetext}
where we have taken into account that $\bar{u}({\bf p}){\bm \gamma} u({\bf p}) = 2{\bf p}$ and the time-dependent term has vanished identically. 
Then we apply the following relation for the helicity states (the upper- and the lower indices are not distinguished here):
\begin{eqnarray}
& \displaystyle 
\bar{u}(p)\,\gamma_j\,\frac{\partial u(p)}{\partial  p_k} - \left (\frac{\partial \bar{u}(p)}{\partial  p_k} \right )\gamma_j\, u(p) = \cr
& \displaystyle = 2i \left (\frac{p_k}{\varepsilon}\, \frac{1}{\varepsilon + m}\, [{\bm \zeta} \times {\bf p}]_j + [{\bm \zeta} \times \hat{{\bf j}}]_k \right ),\, {\bm \zeta} = 2\lambda \hat{\bm z},
\label{mue1}
\end{eqnarray}
where $j,k=1,2,3$; and $\hat{{\bf j}}, |\hat{{\bf j}}| = 1,$ is a unit vector of the $j$-th axis. In the end, we arrive at the following averages:
\begin{eqnarray}
& \displaystyle 
{\bm \mu}_f = \left \langle \frac{1}{(2\varepsilon)^2} \left ({\bm \zeta}(\varepsilon + m) + \frac{{\bf p} ({\bf p {\bm \zeta}})}{\varepsilon + m}\right)\right\rangle 
+ \frac{1}{2} \left \langle {\bm u} \times \frac{\partial \varphi(p)}{\partial {\bm p}}\right \rangle \equiv {\bm \mu}_{s} + {\bm \mu}_{b},
\label{muf12}
\end{eqnarray}
where the definition of a mean value $\langle...\rangle$ is from Eq.(\ref{energymeanexvortex}) and $\varphi(p)$ is a bosonic phase,  
\begin{eqnarray}
& \displaystyle 
\psi (p) \equiv |\psi (p)|\,e^{i\varphi(p)}.
\label{varphi}
\end{eqnarray}
The first term in (\ref{muf12}), ${\bm \mu}_{s}$, describes a spin contribution to the magnetic moment, while ${\bm \mu}_{b}$ represents magnetic moment of a boson. 
It might seem that both the contributions, that of the phase and that of the spin, do not mix, that is, there is no spin-orbit coupling. 
It is indeed the case if the phase $\varphi(p)$ has no singularities (say, for the Airy beams), but for vortex packets the square $|\psi_{\ell}(p)|^2$ still depends on $\ell$ 
and that is why the spin-orbit terms like $\ell{\bm \zeta}$ survive in ${\bm \mu}_{s}$.

Taking then the non-paraxial states (\ref{OAMrelp}), one can approximately evaluate the orbital integral in (\ref{muf12}) 
via the steepest descent method, which yields the following result in the laboratory frame of reference:
\begin{eqnarray}
& \displaystyle 
{\bm \mu}_{b} = \frac{\ell}{2} \left \langle {\bm u} \times \frac{\partial \phi_p}{\partial {\bm p}}\right \rangle = \hat{{\bm z}}\, \ell \left\langle \frac{1}{2\varepsilon}\right\rangle \simeq \hat{{\bm z}}\, \ell\, \frac{1}{2\bar{\varepsilon}} \left (1 - \frac{\sigma^2}{2m^2} \left (|\ell| + \frac{1}{2} + \frac{m^2}{\bar{\varepsilon}^2}\right )\right ),
\label{muf2app}
\end{eqnarray} 
with the correction being negative. Neglecting the latter, we return to the result by Bliokh \textit{et al.}, ${\bm \mu}_{b} = \hat{{\bm z}}\, \ell/2\bar{\varepsilon}$ \cite{Bliokh, Review}. As before, the correction is of the order of $|\ell|\sigma^2/m^2$, even though the integral itself also brings the terms $|\ell|^2\sigma^2/m^2 \gg |\ell|\sigma^2/m^2$.
It is the corresponding $\ell^2$-summand in the expansion of the normalization constant $1/K_{|\ell| + 1}(2m^2/\sigma^2)$ from (\ref{OAMrelp}) that cancels this term,
and this is one of the major reasons why these results cannot be obtained with the simplified paraxial states.

More interestingly, the correction turns out to be \textit{not Lorentz invariant}. This happens because the function to be averaged, $1/2\varepsilon$, is not a component of a tensor 
(cf. Eq.(\ref{energymeanexvortex}) with the energy being a $p^0$).
In the relativistic regime, the non-invariant term $m/\bar{\varepsilon}$ represents a ratio of the particle's de Broglie wavelength, 
$\lambda_{dB} \sim 1/\bar{p}$, to its Compton wavelength $\lambda_c$. The standard interpretation of smallness of this ratio is that the packet's motion becomes quasi-classical 
for higher energies ($\mathcal O(\hbar^2)$-terms can be neglected). However, for vortex beams there are also the terms in (\ref{muf2app}) that are invariant and, 
therefore, do not vanish when $\bar{\varepsilon} \gg m$. In other words, in contrast to the packets with a non-singular phase, the purely quantum corrections influence the vortex packet's motion even in the relativistic case.

%Thus, the non-paraxial corrections are \textit{frame-dependent} for some observables. 
In fact, a pair of dipole moments $({\bf d}, {\bm \mu})$ transforms as a product of an anti-symmetric $4$-tensor and a volume $V$. As the latter is not invariant, %this pair $({\bf d}, {\bm \mu})$ itself does not constitute a $4$-tensor and 
it is $\varepsilon {\bm \mu}$, not ${\bm \mu}$, that transforms as a component of a tensor. %the magnetic moment ${\bm \mu}$ transforms as a product ${\bm \mu} \sim {\bm \mu}_{\text{vec}} V \sim {\bm \mu}_{\text{vec}}/\varepsilon$ where  
Obviously, it is the reason for non-invariance of the correction to the magnetic moment.
 
The calculations for the spin contribution are more challenging and the result is
%\begin{widetext}
\begin{eqnarray}
& \displaystyle 
{\bm \mu}_{s} = \left \langle \frac{1}{(2\varepsilon)^2} \left ({\bm \zeta}(\varepsilon + m) + \frac{{\bf p} ({\bf p {\bm \zeta}})}{\varepsilon + m}\right)\right\rangle \simeq 
{\bm \zeta}\, \frac{1}{2\bar{\varepsilon}} \Bigg (1 - \frac{\sigma^2}{2m^2} \Big [\frac{1}{2} + \frac{3}{2}\frac{m}{\bar{\varepsilon}} + \frac{1}{2}\frac{m^2}{\bar{\varepsilon}^2} - \frac{3}{2}\frac{m^3}{\bar{\varepsilon}^3} - \cr
& \displaystyle - \frac{m}{\bar{\varepsilon} + m} \left (\frac{3}{2} - 2\frac{m^2}{\bar{\varepsilon}^2} - \frac{3}{2}\frac{m^3}{\bar{\varepsilon}^3} \right ) + |\ell| \left(1 + \frac{m}{\bar{\varepsilon}} - \frac{m}{\bar{\varepsilon} + m}\right ) \Big ]\Bigg )
\label{mufsapp}
\end{eqnarray}
%\end{widetext}
where the two last summands represent a parameter $\Delta$ used for characterizing the spin-orbit coupling in \cite{Bliokh, Review},
\begin{eqnarray}
& \displaystyle 
\Delta = \left(1 - \frac{m}{\bar{\varepsilon}}\right) \sin^2\theta_0 \simeq |\ell|\frac{\sigma^2}{m^2} \left(\frac{m}{\bar{\varepsilon}} - \frac{m}{\bar{\varepsilon} + m}\right), 
\label{Delta}
\end{eqnarray}
with the only difference that now it grows with $|\ell|$. 

Then there has appeared a spin-orbit interaction, 
\begin{eqnarray}
& \displaystyle
{\bm \zeta}|\ell|\,\frac{\sigma^2}{m^2} = {\bm \zeta}|\ell|\,\frac{\lambda_c^2}{\sigma_{\perp}^2},
\label{spinorb}
\end{eqnarray}
which is also $|\ell|$ times enhanced compared to the Bessel beam (cf. Eq.(20) in \cite{Bliokh}).
As the total magnetic moment ${\bm \mu}_f$ represents a sum of (\ref{muf2app}) and (\ref{mufsapp}), 
this term will be obscured by the large orbital contribution, which is $\ell$ times stronger.
As a result, although the spin-orbit effects are enhanced for highly twisted electrons, 
the detection of them seems hardly feasible in near future (in accordance with \cite{Boxem}).
Note that separation of the magnetic moment into the orbital part and the spin one is unique, as ${\bm r}$ in the left-hand side of (\ref{muf}) is not an operator (see also \cite{Bliokh17}).

Finally, similarly to (\ref{muf}), one can also calculate an electron's electric dipole moment,
\begin{eqnarray}
& \displaystyle 
{\bm d}_f = \int d^3r\, {\bm r}\, j^0 = \left \langle {\bm u} t - \frac{\partial \varphi (p)}{\partial {\bm p}} + \frac{{\bm p}\times {\bm \zeta}}{2\varepsilon (\varepsilon + m)} \right \rangle 
\label{df}
\end{eqnarray}
where we have used that 
$$
u^{\dagger}(p)\frac{\partial u(p)}{\partial {\bm p}} - \text{c.c.} = \frac{2i}{\varepsilon + m}\, {\bm \zeta}\times{\bm p}.
$$
Due to azimuthal symmetry of the vortex states, two last terms in (\ref{df}) vanish and so ${\bm d}_f = \langle {\bm u}\rangle t = 0$ at $t=0$. 
One can also define a mean path of the electron packet via its dipole moment as $\langle {\bm r} \rangle := {\bm d}_f/\int d^3r j^0 = \langle {\bm u}\rangle t$
where the mean velocity $\langle {\bm u}\rangle = \bar{{\bm u}} (1 + \mathcal O(|\ell|\sigma^2/m^2))$ also acquires the frame-dependent corrections,
as it is ${\bm u}\,\varepsilon/m$ that represents a spatial component of a $4$-velocity, not ${\bm u}$.

\section{Summary} 

Although the paraxial approximation is applicable for available vortex electrons in the majority of cases,
the corrections to it are crucial for many potential applications and they can be accurately estimated only when using well-localized wave packets described in a Lorentz invariant way.
We have presented such non-paraxial generalizations of the LG beams and calculated the corresponding corrections. 
As it turns out, they are linearly enhanced for large OAM $|\ell| \gg 1$ and, therefore, can be several orders of magnitude larger than those for the customary Gaussian packets.

For available beams with $\sigma_{\perp} < 1\,\text{nm}$ and $|\ell| > 10^3$, this allows one to probe for the first time the non-paraxial effects in particle physics,
which are analogous to the recently studied quantum coherence phenomena in atomic collisions \cite{Sarkadi, Sch}.
It is the high OAM that can help to overcome the overall $\lambda_c/a = 1/137$ attenuation of these effects compared to the atomic scale.
% In order to reach the hadronic scale of $\langle p_{\perp}\rangle \sim 10$ MeV for the electrons focused to $\sigma_{\perp} \sim 0.1$ nm, 
% one would need to get the OAM as high as $\ell \sim 10^8$.

One of these effects is a correction to the plane-wave cross section in $e^-e^-$ or $e^-\gamma$ high-energy scattering with vortex electrons and photons. 
With current technology, these corrections can reach the relative values of $10^{-4} - 10^{-3} \gtrsim \alpha_{em}^2$, that is, become comparable with the two-loop QED contribution.
Thus, highly twisted relativistic beams can become a new tool in the high-energy physics. The corresponding calculations can be performed analogously to \cite{JHEP}
by using the quantum states described in this paper.

I am grateful to V.~Bagrov, I.~Ginzburg, I.~Ivanov, D.~Naumov, V.~Serbo, A.~Zhevlakov, and, especially, to P.~Kazinski for many useful discussions and criticism. 
This work is supported by the Russian Science Foundation (project No.\,17-72-20013).

\end{document}